\documentclass[11pt,a4paper]{article}
\pdfoutput=1
\usepackage{jcappub}

\usepackage{bm}
\usepackage{graphicx}
\usepackage{caption}
\usepackage{subcaption}

\newcommand{\neff}{N_{\textrm{eff}}}

\newcommand{\meff}{m_{\textrm{s}}^{\textrm{eff}}}

\newcommand{\Hou}{\ensuremath{\text{Km s}^{-1}\text{ Mpc}^{-1}}}
\newcommand{\lcdm}{$\Lambda$CDM}
\newcommand{\lcdmnus}{$\Lambda$CDM $+\,\neff \,+\,m_s$}

\newcommand{\nua}[1]{\ensuremath{\rlap{\kern-2.5pt\ensuremath{\overset{\scriptscriptstyle(-)}{\phantom{\nu}}}}{\ensuremath{{\nu}_{#1}}}}}

\renewcommand{\nua}[1]{\nu_{#1}~(\bar{\nu}_{#1})}

\begin{document}

%%%%%%%%%%%%%%%%%%%%%%%%%%%%%%%%%%%%%%%%%%%%%%%%%%%%%%%%%%%%%%%%%%%%%%
% Frontpage %%%%%%%%%%%%%%%%%%%%%%%%%%%%%%%%%%%%%%%%%%%%%%%%%%%%%%%%%%
%%%%%%%%%%%%%%%%%%%%%%%%%%%%%%%%%%%%%%%%%%%%%%%%%%%%%%%%%%%%%%%%%%%%%%

%\subheader{\hfill Preprint ....}

\title{Pseudoscalar - sterile neutrino interactions: reconciling the cosmos with neutrino oscillations}

\author[a]{Maria Archidiacono}
\emailAdd{archidiacono@physik.rwth-aachen.de}

\author[b,c]{, Stefano Gariazzo}
\emailAdd{gariazzo@to.infn.it}

\author[c]{, Carlo Giunti}
\emailAdd{giunti@to.infn.it}

\author[d]{, Steen Hannestad}
\emailAdd{steen@aias.au.dk}

\author[e]{, Rasmus Hansen}
\emailAdd{rasmus@mpi-hd.mpg.de}

\author[f]{, Marco Laveder}
\emailAdd{laveder@pd.infn.it}

\author[g]{, Thomas Tram}
\emailAdd{thomas.tram@port.ac.uk}

\affiliation[a]{Institute for Theoretical Particle Physics and Cosmology (TTK), RWTH Aachen University, D-52056 Aachen, Germany.}
\affiliation[b]{Department of Physics, University of Torino, Via P. Giuria 1, I--10125 Torino, Italy}
\affiliation[c]{INFN, Sezione di Torino, Via P. Giuria 1, I--10125 Torino, Italy}
\affiliation[d]{Department of Physics and Astronomy, Aarhus University, 8000 Aarhus C, Denmark}
\affiliation[e]{Max-Planck-Institut f\"ur Kernphysik, Saupfercheckweg 1, 69117 Heidelberg, Germany}
\affiliation[f]{Dipartimento di Fisica e Astronomia ``G. Galilei'', Universit\`a di Padova, and INFN, Sezione di Padova, Via F. Marzolo 8, I--35131 Padova, Italy}
\affiliation[g]{Institute of Gravitation and Cosmology, University of Portsmouth, Dennis Sciama Building, Burnaby Road, Portsmouth, PO1 3FX, United Kingdom}

\abstract{
The Short BaseLine (SBL) neutrino oscillation anomalies hint
at the presence of a sterile neutrino with a mass of around 1 eV.
However, such a neutrino is incompatible
with cosmological data,
in particular observations of the Cosmic Microwave Background (CMB) anisotropies. 
However, this conclusion can change by invoking new physics. One possibility is 
to introduce a secret interaction in the sterile neutrino sector mediated by a light pseudoscalar. In this pseudoscalar model, CMB data prefer a sterile
neutrino mass that is fully compatible with the mass ranges suggested by SBL anomalies.
In addition, this model predicts a value of the Hubble parameter which is completely consistent
with local measurements.
}

\maketitle

\section{Introduction}

One of the long-standing open issues of modern physics is the tension between cosmological constraints on the effective number of relativistic degrees of freedom $\neff$ and on the neutrino mass sum and the hints for sterile neutrinos from oscillation experiment~\cite{Abazajian:2012ys,Kopp:2013vaa, Gariazzo:2015rra, Gonzalez-Garcia:2015qrr}.
Indeed, precise measurements of the primordial abundances at the time of Big Bang Nucleosynthesis rule out a fourth sterile neutrino with high significance~\cite{Cooke:2013cba,Cyburt:2015mya}.
Moreover, the tension with Planck Cosmic Microwave Background (CMB) data~\cite{Ade:2015xua} is twofold: on the one hand CMB data are fully consistent with the standard cosmological value of $\neff$ ($\neff=3.04\pm0.18$, Planck TT, TE, EE + lowP + BAO), 
on the other constraints on the neutrino mass sum are tight ($\Sigma m_\nu<0.17$ eV at 95\% c.l., Planck TT, TE, EE + lowP + BAO).
Invoking the tension between CMB and either local measurements of the Hubble constant $H_0$~\cite{Riess:2016jrr} or the lensing measurements of the clumpiness of the local Universe (i.e.\ $\sigma_8 \left( \Omega_m/\Omega_{m0} \right)^\gamma$) does not represent an escape route.
Concerning the latter discrepancy, the situation is still unclear. The latest results from the Dark Energy Survey~\cite{Kwan:2016mcy} alleviated the tension found by CFHTLenS~\cite{Kilbinger:2012qz}, while the recently published intermediate results of KiDS~\cite{Hildebrandt:2016iqg} find a $2.3\sigma$ tension with Planck, in the \lcdm~model.
Indeed the joint analysis performed by Planck in the context of an extended cosmological model simultaneously constraining $\neff$ and $m_s$%
\footnote{More precisely the constraint is on the so-called effective sterile neutrino mass, which for additional thermally distributed degrees of freedom $N_s$ is related to the physical mass through $\meff=(N_s)^{3/4}m_s$.},
namely \lcdmnus\ model, shows that the global $\Delta\chi^2$ does not improve significantly, and the $\neff$ and $m_s$ constraints remain far away from the oscillation preferred regions, considering that $\Delta m^2 \sim 1$ eV and that the large mixing angle would lead to $\neff \sim 4$~\cite{Hannestad:2012ky}.
The only way to reconcile eV sterile neutrinos in cosmology is by delaying their production and, thus, suppressing their thermalisation in the early Universe~\cite{Hannestad:2012ky,Saviano:2013ktj}. Thus, this highly debated tension between cosmology and neutrino oscillations may point to physics beyond the standard model. 

In this paper we show how eV sterile neutrinos, highly disfavoured in the standard thermalised case, can be naturally accommodated within the pseudoscalar interaction model. 
The paper is organised as follows:
in section \ref{sec:sbl} we report the latest global-fit of laboratory experiments;
in section~\ref{sec:cosmological} we derive cosmological constraints on light sterile neutrinos in both the standard and the pseudoscalar model;
in section~\ref{sec:joint} we present the results of the combined analyses of cosmological and oscillation data, within the context of the pseudoscalar model;
finally section~\ref{sec:discussion} summarises our findings.

\section{Short-baseline neutrino oscillations}
\label{sec:sbl}

There are three anomalies found in short-baseline neutrino oscillation experiments
which cannot be explained in the standard framework of
three-neutrino mixing
(see the recent reviews in refs.~\cite{Gariazzo:2015rra,Gonzalez-Garcia:2015qrr}):

\begin{enumerate}

\item
The measurement of
$\bar\nu_{\mu}\to\bar\nu_{e}$
transitions in the LSND short-baseline neutrino oscillation experiment
\cite{Athanassopoulos:1995iw,Aguilar:2001ty}.

\item
The short-baseline disappearance of $\nu_{e}$
\cite{Laveder:2007zz,Giunti:2006bj,Giunti:2010zu,Giunti:2012tn}
measured in the
Gallium radioactive source experiments
GALLEX
\cite{Kaether:2010ag}
and
SAGE
\cite{Abdurashitov:2009tn}.

\item
The deficit of the rate of $\bar\nu_{e}$ events
\cite{Mention:2011rk}
observed in several
short-baseline reactor neutrino experiments
in comparison with that expected from the calculation of
the reactor neutrino fluxes
\cite{Mueller:2011nm,Huber:2011wv,Abazajian:2012ys}.

\end{enumerate}

These anomalies can be explained by neutrino oscillations if,
besides the standard three massive neutrinos
$\nu_{1}$,
$\nu_{2}$,
$\nu_{3}$
which are mainly mixed with the three standard active neutrinos
$\nu_{e}$,
$\nu_{\mu}$,
$\nu_{\tau}$,
there is at least one new massive neutrino
$\nu_{4}$
which is mostly sterile and has a mass at the eV scale.
In this so-called 3+1 neutrino mixing scheme,
the new massive neutrino generates a new short-baseline squared-mass difference
\begin{equation}
\Delta{m}^2_{\text{SBL}}
=
\Delta{m}^2_{41}
\simeq
\Delta{m}^2_{42}
\simeq
\Delta{m}^2_{43}
\gtrsim
1 \, \text{eV}^2
,
\label{sbl}
\end{equation}
with
$\Delta{m}^2_{kj} \equiv m_{k}^2 - m_{j}^2$,
which is much larger than the standard solar squared-mass difference
$
\Delta{m}^2_{\text{SOL}}
=
\Delta{m}^2_{21}
\simeq
7.5 \times 10^{-5} \, \text{eV}^2
$
and the standard atmospheric squared-mass difference
$
\Delta{m}^2_{\text{ATM}}
=
|\Delta{m}^2_{31}|
\simeq
|\Delta{m}^2_{32}|
\simeq
2.3 \times 10^{-3} \, \text{eV}^2
$
(see refs.~\cite{Forero:2014bxa,Gonzalez-Garcia:2014bfa,Capozzi:2016rtj}).

The short-baseline anomalies are also compatible with a 1+3
neutrino mixing scheme in which the squared-mass differences
in Eq.~(\ref{sbl}) have opposite sign,
but cosmological data
exclude the existence of three standard massive neutrinos at the eV scale
\cite{Ade:2015xua}.
Taking into account that the current cosmological limit on the sum of the masses
of the three standard neutrinos is well below the eV scale
and the experimental bounds on
neutrinoless double-$\beta$ decay\footnote{
Assuming that massive neutrinos are Majorana particles
(see refs.~\cite{Bilenky:2014uka,Dell'Oro:2016dbc}).
},
which imply that the scale of the masses of the three standard neutrinos
is much smaller than 1 eV
\cite{KamLAND-Zen:2016pfg},
we consider the 3+1 scheme with
$m_{1},m_{2},m_{3} \ll m_{4}$.
In this case the approximation
$m_{4}^2 \simeq \Delta{m}^2_{41} = \Delta{m}^2_{\text{SBL}}$
allows us to compare and combine the results of the analysis of
cosmological data,
which is sensitive to $m_{4}$,
and the results of the analysis of the data
of short-baseline neutrino oscillation experiments.
Following a common convention,
in the rest of the paper we will often
denote $m_{4}$ as $m_{s}$,
since $\nu_{4}$ is mostly sterile.

\begin{figure*}
\begin{tabular}{cc}
\includegraphics*[width=0.48\linewidth]{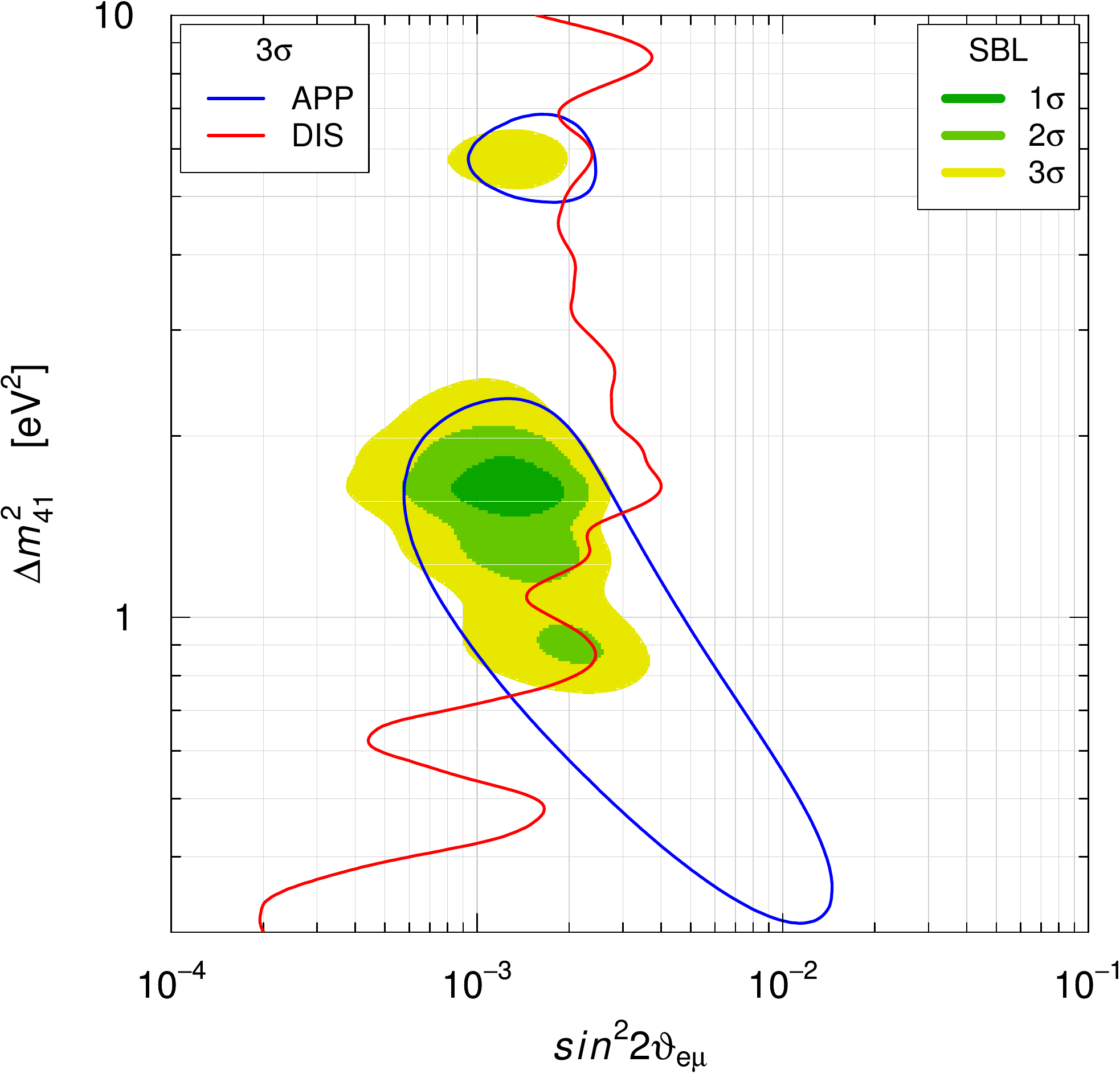}
&
\includegraphics*[width=0.48\linewidth]{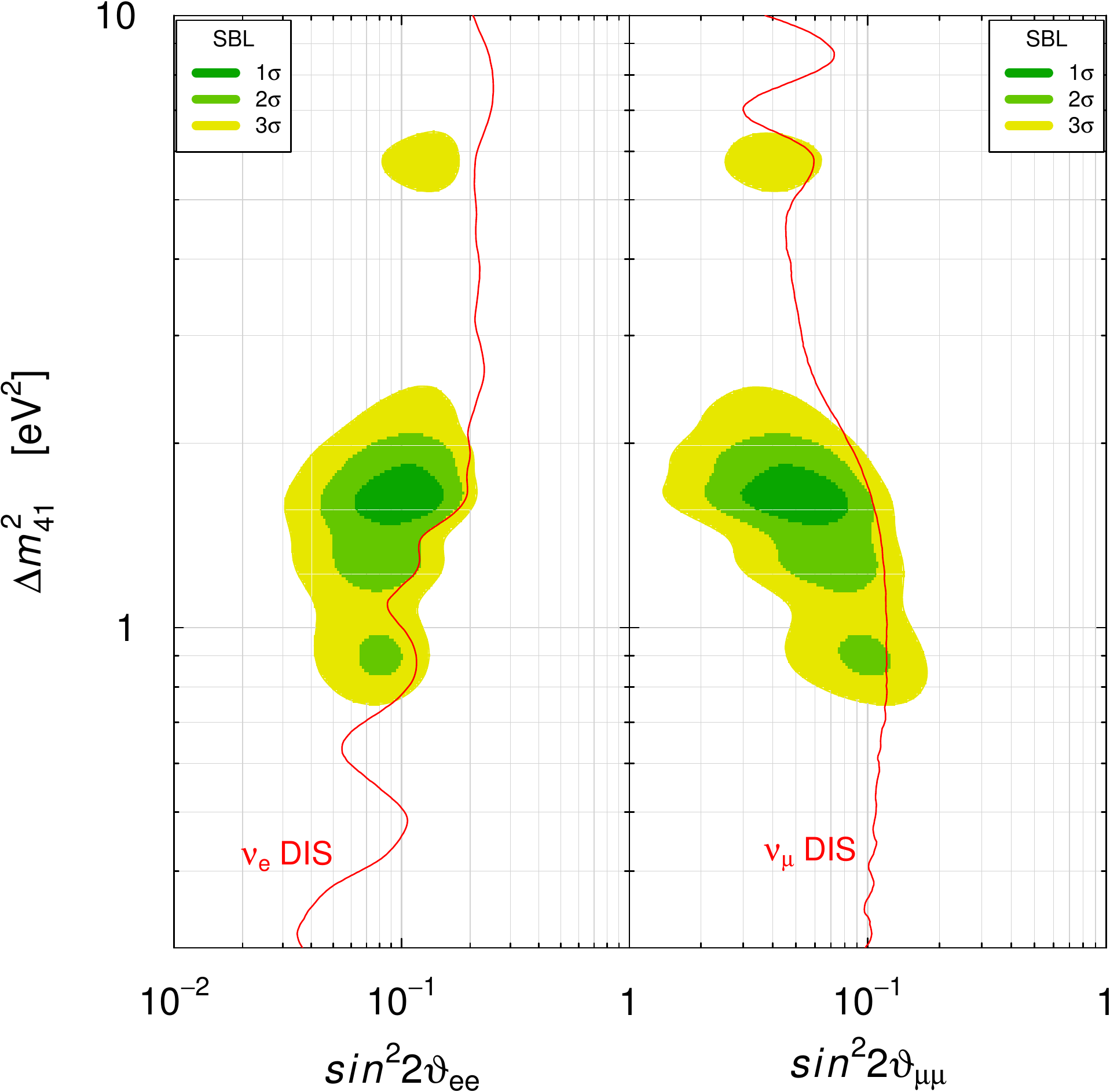}
\end{tabular}
\caption{ \label{fig:sbl}
Regions in the planes of the effective amplitudes
$\sin^22\vartheta_{e\mu}$,
$\sin^22\vartheta_{ee}$ and
$\sin^22\vartheta_{\mu\mu}$
versus
$\Delta{m}^2_{41}$
which are allowed by the Bayesian global fit of short-baseline (SBL)
neutrino oscillation data.
Also shown are
the $3\sigma$ bounds obtained from the separate fits of appearance
(APP; the regions inside the two blue contours are allowed)
and disappearance
(DIS; the regions on the left of the red lines are allowed) data.
$1\sigma$,
$2\sigma$ and
$3\sigma$
correspond, respectively,
to
68.27\%,
95.45\% and
99.73\%
posterior probability.
}
\end{figure*}

We have performed a Bayesian global fit
to the short-baseline neutrino oscillation data
considered in ref.~\cite{Gariazzo:2015rra}.
In the framework of 3+1 neutrino mixing,
the effective probability of
$\nu_\alpha \to \nu_\beta$ and
$\bar{\nu}_\alpha \to \bar{\nu}_\beta$
transitions in short-baseline experiments is given by
\cite{Bilenky:1996rw}
\begin{equation}
P_{\nu_\alpha \to \nu_\beta} = 
P_{\bar{\nu}_\alpha \to \bar{\nu}_\beta}
=
\delta_{\alpha\beta}
-
4 |U_{\alpha4}|^2 \left( \delta_{\alpha\beta} - |U_{\beta4}|^2 \right)
\sin^2\!\left( \dfrac{\Delta{m}^2_{41}L}{4E_{\nu}} \right)
,
\label{pab}
\end{equation}
where $U$ is the $4\times4$ mixing matrix
in the mixing relation of the left-handed neutrino fields:
\begin{equation}
\nu_{\alpha L}
=
\sum_{k=1}^{4}
U_{\alpha k} \nu_{k L}
\qquad
(\alpha=e,\mu,\tau,s)
.
\label{mixing}
\end{equation}
We considered the following short-baseline data sets:
\begin{itemize}
\item $\nu_{\mu}\to\nu_{e}$ and $\bar\nu_{\mu}\to\bar\nu_{e}$
transitions,
with amplitude
$\sin^22\vartheta_{e\mu} = 4 |U_{e4}|^2 |U_{\mu4}|^2$,
\item $\nu_{e}$ and $\bar\nu_{e}$ disappearance,
with amplitude
$\sin^22\vartheta_{ee} = 4 |U_{e4}|^2 \left( 1 - |U_{e4}|^2 \right)$,
\item $\nu_{\mu}$ and $\bar\nu_{\mu}$ disappearance\footnote{
We did not consider the recent results of the IceCube
\cite{TheIceCube:2016oqi}
and MINOS
\cite{Timmons:2016hvv}
experiments,
which constrain $\sin^22\vartheta_{\mu\mu}$
for
$\Delta{m}^2_{41} \lesssim 1 \, \text{eV}^2$,
because we do not have sufficient information for the data analysis.
},
with amplitude
$\sin^22\vartheta_{\mu\mu} = 4 |U_{\mu4}|^2 \left( 1 - |U_{\mu4}|^2 \right)$.
\end{itemize}
Although there are three effective angles which determine the oscillations,
$\vartheta_{ee}$,
$\vartheta_{\mu\mu}$ and
$\vartheta_{e\mu}$,
their values depend only on the absolute values of two elements of the
mixing matrix,
$U_{e4}$ and $U_{\mu4}$.

The results of the Bayesian global fit of short-baseline
neutrino oscillation data is shown in figure~\ref{fig:sbl},
where we have plotted the allowed regions in the planes of the effective amplitudes
$\sin^22\vartheta_{e\mu}$,
$\sin^22\vartheta_{ee}$ and
$\sin^22\vartheta_{\mu\mu}$
versus
$\Delta{m}^2_{41}$.
These allowed regions can be compared with those in figure~4
of ref.~\cite{Gariazzo:2015rra},
which have been obtained with a $\chi^2$ analysis.
One can see that the Bayesian allowed regions are slightly larger than the
$\chi^2$-allowed regions and there is a Bayesian region allowed at $3\sigma$
at
$\Delta{m}^2_{41} \approx 6 \, \text{eV}^2$
which does not appear in the $\chi^2$ analysis.
Figure~\ref{fig:sbl} shows also the $3\sigma$ bounds obtained from the
separate fits of
appearance and disappearance data.
From the left panel one can see that
there is an appearance-disappearance tension
(see ref.~\cite{Gariazzo:2015rra})
and
the large-$\sin^22\vartheta_{e\mu}$
part of the region allowed by the appearance data
is excluded by the disappearance data.

\section{Cosmology and light sterile neutrinos}
\label{sec:cosmological}

\subsection{Standard light sterile neutrinos}
\label{sec:lsn}
Cosmology can constrain the properties of the additional neutrino~\cite{Lesgourgues:2014zoa},
through its contribution to the effective number of relativistic species $N_\text{eff}$ and its mass $m_s\simeq m_4$ (see section~\ref{sec:sbl}).
Several analyses in this direction were performed in the past, see e.g.~refs.~\cite{Archidiacono:2012ri,Archidiacono:2013xxa,Kristiansen:2013mza,Ade:2013zuv,Hamann:2013iba,Gariazzo:2013gua,Archidiacono:2014apa,Bergstrom:2014fqa,Ade:2015xua,Gariazzo:2015rra}.
Here we update the cosmological constraints on neutrino related parameters
using the most recent cosmological data. The cosmological model is described by the following set of parameters: 
\begin{equation}\label{eq:cosmoParSpace}
\left\lbrace \omega_{\rm b}, \, \omega_{\rm cdm},\,\theta,\,\tau,\,n_s,\,A_s,\, \neff, \, m_s \right\rbrace
\end{equation}
where $\omega_{\rm cdm} \equiv \Omega_{\rm cdm} h^2$ and $\omega_{\rm b}
\equiv \Omega_{\rm b} h^2$ are the present-day physical CDM and baryon densities
respectively,
$\theta$ is the size of the sound horizon at recombination,
$\tau$ is the optical depth to reionisation,
and $A_{\rm s}$ and $n_s$ denote respectively the amplitude and spectral index of the initial power spectrum of the scalar fluctuations.
The $\neff$ parameter encodes the contribution of the active and sterile neutrinos to the radiation energy density in the early Universe:
$\neff = 3.046+\Delta\neff$, where $\Delta \neff$ accounts for the sterile neutrino degrees of freedom and 3.046 is the standard contribution of the active neutrinos~\cite{Mangano:2005cc}.

We consider two distinct models for the standard sterile neutrino. In the first one 
we consider an extension of the \lcdm\ model
where we introduce a sterile neutrino with mass $m_s$
contributing with $\Delta\neff$ to the value of $\neff$.
Both $\Delta\neff$ and $m_s$ are allowed to vary and we will refer to this model as ``\lcdmnus''.
In the second one we will allow the mass to vary, but we will assume that the sterile neutrino is fully thermalised, i.e.\ $\Delta\neff=1$. 
This model is motivated by the SBL constraints on the mixing angles which would lead to $\Delta\neff \simeq1$~\cite{Hannestad:2012ky} for a wide range of $m_s$. We will denote this model by  ``\lcdm+$1\nu_s$''.

We sample the parameter space~\eqref{eq:cosmoParSpace} using the Markov Chain Monte Carlo sampler CosmoMC~\cite{Lewis:2002ah} and data from both the CMB and local measurements.
We use the Planck 2015 data release~\cite{Aghanim:2015xee} for the CMB data and include low multipole polarisation and high multipole temperature auto-correlation data as our base-line dataset (denoted by ``TT'' for the sake of brevity)
and, in addition, in combination with high $\ell$ E-mode polarisation auto-correlation and temperature cross correlation (TTTEEE).
Local Universe probes include the prior on the Hubble constant from direct measurements, $H_0=73.00\pm1.75\,\,\Hou$~\cite{Riess:2016jrr} (HST)
and Baryonic Acoustic Oscillations (BAO) from a list of different experiments, that probe different redshifts: 6dFGS~\cite{Beutler:2011hx}, SDSS-MGS~\cite{Ross:2014qpa}, and from the LOWZ~\cite{Anderson:2012sa} and CMASS~\cite{Anderson:2013zyy} DR11 results from the SDSS BOSS experiment.

\begin{figure*}[t]
\centering
\includegraphics*[width=0.6\linewidth]{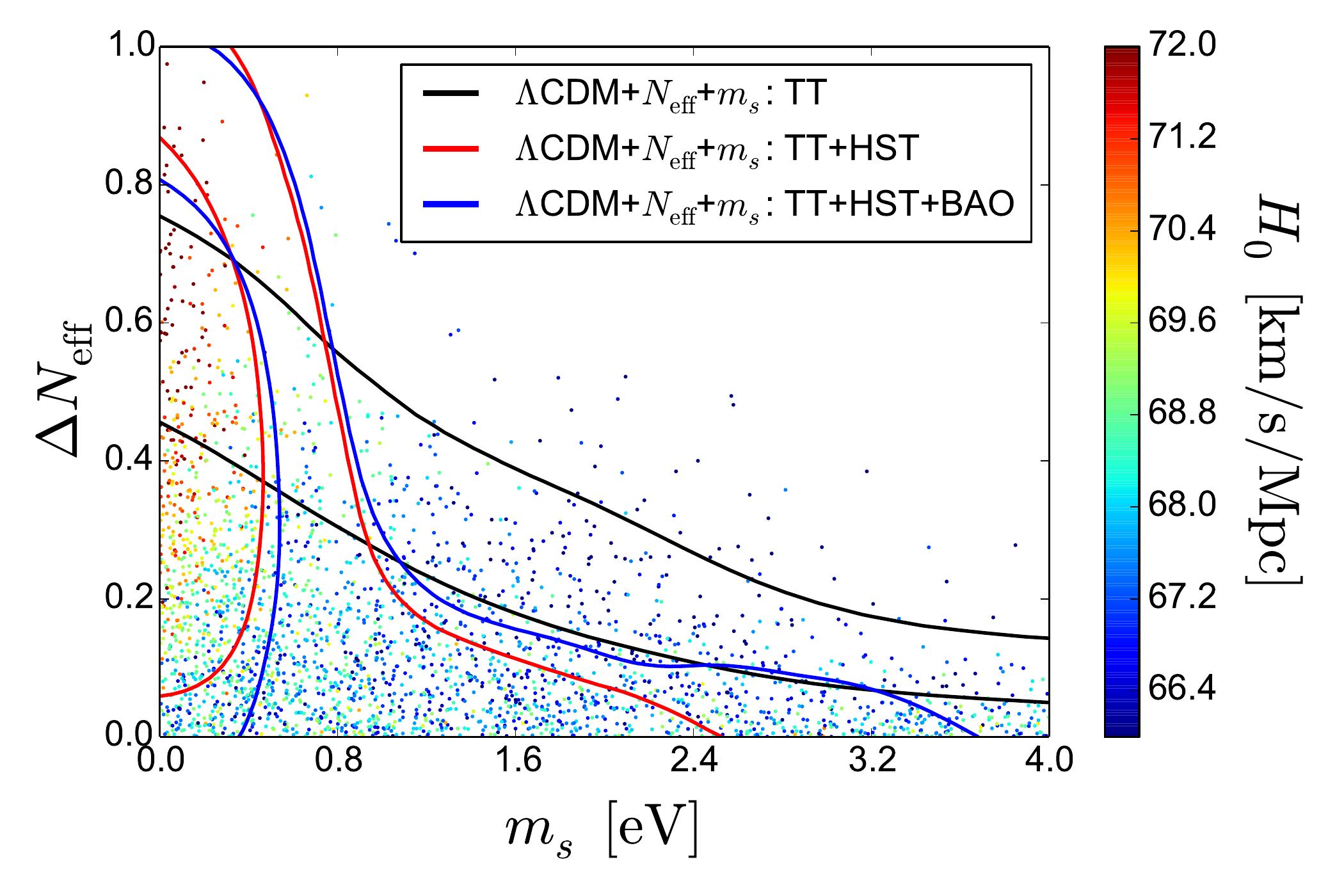}
\caption{ \label{fig:msNnuPost}
Marginalised constraints in the $m_s$--$\Delta\neff$ plane,
obtained from the analyses of the cosmological data
in the context of the \lcdmnus\ model,
at 1$\sigma$ and 2$\sigma$ confidence level.
The excluded regions are on the right of each line.
The points obtained with the TT dataset are colour-coded by their $H_0$ value.
}
\end{figure*}

The cosmological constraints on the neutrino properties in the \lcdmnus\ model
are summarised in figure~\ref{fig:msNnuPost},
where we plot the 2D marginalised constraints in the $m_s$--$\Delta \neff$ plane
as obtained from some selected combinations of cosmological data.
The points are colour-coded by their $H_0$ value.
The results are in full agreement with those presented by the Planck collaboration~\cite{Ade:2013zuv,Ade:2015xua},
even if the parameterization they adopt involves the effective mass $\meff$
instead of the sterile neutrino physical mass $m_s$.
We can easily see that a sterile neutrino with $m_s\simeq1$~eV
and contributing with $\Delta\neff \simeq1$
is strongly disfavoured by all the data combinations.
As stated in ref.~\cite{Riess:2016jrr},
the possibility to increase $\neff$ improves the compatibility
between the cosmological estimates and the local measurement of $H_0$,
but this is true only for a sterile neutrino much lighter than 1~eV.
As we can see in table~\ref{tab:lsnBounds}, indeed, the marginalised
constraints on $H_0$ obtained from the \lcdmnus\ model are
slightly more compatible with the local value determined by HST,
but the tension remains strong.

\begin{table*}[t]
\begin{center}
\resizebox{1\textwidth}{!}{
\begin{tabular}{lccccc}
\hline
\hline
 Parameter                & TT                     &TT+HST                   &TT+BAO                  &TT+HST+BAO               &TTTEEE\\[0.1cm]

\hline\\[0.001cm]
$H_0 \, [{\rm km/s/Mpc}]$ & $68.0\,^{+1.0}_{-1.5}$ & $70.7\,^{+1.7}_{-2.0}$  & $68.3\,^{+0.6}_{-1.0}$ & $69.8\,^{+1.2}_{-1.5}$  & $67.0\,^{+0.7}_{-0.8}$ \\[0.1cm]\hline
$\neff$                   & $<3.53$                & $<3.88$                 & $<3.49$                & $<3.84$                 & $<3.36$      \\[0.1cm]

\hline
\hline
\end{tabular}
}
\end{center}
\caption{Constraints on $H_0$ and $\neff$ in the \lcdmnus\ model.
Marginalised constraints are given at $1\sigma$,
while upper bounds are given at $2\sigma$.}
\label{tab:lsnBounds}
\end{table*}

We will now consider the SBL motivated model \lcdm+$1\nu_s$ where $\Delta\neff=1$.
In this case the high value of $\neff$ prevents large values of $m_s$,
and the SBL preferred region is excluded at more than 2$\sigma$
in the most favorable case.
Indeed the 95\% C.L.\ upper bounds obtained are:
$m_s<0.66$~eV for CMB (TT) alone,
$m_s<0.55$~eV including the prior on the Hubble parameter (TT+HST),
and $m_s<0.53$~eV using the most complete data combination
that we adopt here (TT+HST+BAO).
Given that the $m_s\simeq1$~eV neutrino is firmly excluded,
we do not make a combined fit of the cosmological
and SBL data with a fixed $\Delta\neff=1$.
If the existence of the sterile neutrino is confirmed
by future experiments like SOX~\cite{Borexino:2013xxa},
we will be forced to introduce some new physics.
In the next section we will show that the pseudoscalar model is a natural candidate in this case.

\subsection{The pseudoscalar model}

Refs.~\cite{Archidiacono:2014nda,Archidiacono:2015oma} have shown that hidden neutrino interactions, confined in the sterile sector and mediated by a light $<0.1$ eV pseudoscalar of a new U(1) broken symmetry can solve the tension and reconcile eV sterile neutrinos with cosmology.
The model is described by the Lagrangian
\begin{equation}
\mathcal{L} \sim g_s \phi \bar{\nu}_4 \gamma_5 \nu_4,
\end{equation}
where $g_s$ is the coupling and $\phi$ is the pseudoscalar. The phenomenologically success of the model relies on two things. First, $\nu_4$ must annihilate into $\phi$ at late time to avoid the mass bound from large scale structure. For this to work $\phi$ must not only be lighter than the fourth mass eigenstate but also light enough to avoid the mass bound itself. Thus we consider $m_\phi \lesssim 0.1~\text{eV}$ as an upper bound. Second, the coupling $g_s$ should be large enough to prevent full thermalisation of the sterile neutrino. Numerical studies~\cite{Archidiacono:2014nda} show that the transition from full to zero thermalisation happens in the interval $10^{-6} \lesssim g_s \lesssim 10^{-5}$, so we will assume $g_s$ to be in that range.  

The secret coupling has to be confined in the sterile sector so that its effect on active neutrinos is Yukawa suppressed: the Universe does not end up being a ``neutrinoless" Universe~\cite{Beacom:2004yd} and active neutrinos remain free-streaming, as indicated by current data \citep{Cyr-Racine:2013jua,Archidiacono:2013dua}.

The coupling of the mediator is not constrained by fifth force experiments, because the pseudoscalar couples only to the spin and the medium is globally unposarised.  
The IceCube constraints on secret interactions discussed in
refs.~\cite{Ioka:2014kca,Cherry:2014xra,Ng:2014pca,Cherry:2016jol}
do not apply either,
because the secret interaction concerns only the massive neutrino $\nu_4$.
If the astrophysical PeV neutrinos detected by IceCube \cite{Aartsen:2014gkd}
were produced as active flavor neutrinos
they had only a small component of $\nu_4$
and only this small component can in principle be depleted
through the secret interaction by scattering on the pseudoscalar $\phi$ 
(the $\nu_4$ background annihilates away as soon as the temperature drops below 
the $\nu_4$ mass).
However, the cross section for this interaction is exceedingly small since
$\sigma \sim g_{s}^4/s$,
where $s = 2 E T_\phi$ and $E$ is the energy of the astrophysical neutrino.

The only non-cosmological bound on the pseudoscalar coupling to neutrinos is related to the supernova energy loss argument of ref.~\cite{Farzan:2002wx} which implies $g_s \lesssim 10^{-4}$.
Finally the new mediator might also couple to dark matter $\chi$ and the induced $\chi - \nu_s$ scattering\footnote{The same effect can arise from dark matter scattering off of a different dark radiation component rather than sterile neutrinos, such as dark photons or dark gluons~\cite{Lesgourgues:2015wza}.} can provide a solution to the small scale structure problems of $\Lambda$CDM (see refs.~\cite{Bringmann:2013vra,Dasgupta:2013zpn,Chu:2014lja}).

Since pair production only brings the mediator $\phi$ into thermal equilibrium at very late times ($T \ll$ MeV) for the values of $g_s$ considered here there should essentially be no pre-existing population of these particles. Of course $\phi$ could potentially be produced by direct inflaton decay in the very early universe. However, such a population would have been strongly diluted by subsequent entropy production and thus completely negligible.

When neutrinos start oscillating, the MSW-like potential induced by the new interactions with $g_s \gtrsim 10^{-6}$ suppresses the sterile neutrino production until after the collisional decoupling of active neutrinos ($T\sim1$ MeV): when sterile neutrinos are later produced through oscillations with the active neutrinos, the latter are not able to thermalise with the plasma.
This partial thermalisation can also be achieved by means of a large lepton asymmetry~\cite{Hannestad:2012ky,Saviano:2013ktj} or secret interactions mediated by a massive vector boson~\cite{Hannestad:2013ana,Dasgupta:2013zpn,Mirizzi:2014ama,Saviano:2014esa,Forastieri:2015paa,Chu:2015ipa}.

Here it should be noted that~\cite{Cherry:2016jol} pointed out that scatterings mediated by the pseudoscalar can in principle be important for thermalising sterile neutrinos. The reason is that the scattering rate diverges as $1/p^2$ for small momentum transfer as long as $p > m$, where $m$ is the mass of the mediator \footnote{The model studied in~\cite{Cherry:2016jol} is based on a vector mediator. However, the infrared divergence does not depend on this and also occurs in models with scalar or pseudoscalar mediators.}. However, even though the momentum transfer becomes small, the interaction still provides a flavour measurement and thus contributes to the thermalisation of the fourth mass state.
For a model like the one presented here this at first sight leads to a gigantic increase in the thermalisation rate because the mediator mass is very small.
However, the divergence only occurs in vacuum. In a medium which is charge neutral (i.e.\ not spin polarised in the case of a pseudoscalar interaction), Debye screening regulates the divergence through an effective screening mass, $m_D$.
While a quantitative calculation of the screening mass is very involved, we note that the usual Debye mass from QED is $\sim eT$. In comparison, the screening mass is order $T$ in a nuclear medium with interactions through pion exchange (see e.g.~\cite{Florkowski:1993bq}). None of these two examples are exactly equivalent to a gas of sterile neutrinos and pseudoscalars, but they should give a lower and upper estimate for the screening mass. Compared to QED, the screening will be more efficient since the pseudoscalar potential falls of as $1/r^3$ rather than $1/r^2$. On the other hand, the composite nature of the mesons plays an important role for their screening mass, and in the pseudoscalar model we might not expect the screening to be as efficient. With a screening mass around $T$, the rate of flavour measurement through scattering becomes comparable to the one from annihilation/pair-production which is $\sim n_{\nu_s} g^4/T^2$ and completely negligible compared to the standard model flavour measurement rate because of the tiny value of $g^4$. On the other hand, if the screening mass lies closer to the QED value of $gT$, the scattering processes dominate and this could lead to additional thermalisation. Since a full calculation of the screening mass is beyond the scope of this paper and we expect the results only to be modified moderately, we adopt the agnostic view that it will just be another process contributing to $\neff$.

The value of $T$ should represent the temperature of the fourth mass state, i.e.\ it only becomes important once thermalisation begins in earnest. This means that the divergence of the scattering rate can in principle be important, but only provided that the effective temperature of the fourth mass state is very small. Consequently, the thermalisation through scattering is self-regulating and most likely switches off long before complete thermalisation occurs. The most reasonable expectation is thus that there is little or no sterile neutrino thermalisation at high temperature.

However, the crucial point of the pseudoscalar model is related to its late time phenomenology: indeed, the collisional term is inversely proportional to the temperature, so it becomes relevant (i.e.\ $\Gamma_{\rm coll}>H_{\rm Hubble\,rate}$) at late times.
Provided that $g_s \gtrsim 10^{-6}$, as required by the early Universe constraints, the pseudoscalar - sterile neutrino re-coupling occurs before the non-relativistic transition, which, for a $~1$ eV neutrino, is around the time of photon decoupling.
The sterile neutrino annihilation will transfer energy to the pseudoscalar, so the temperature of the fluid will decrease less rapidly than for standard neutrinos ($T \propto a^{-1}$) and the contribution of the fluid to the energy density of relativistic species will be larger than the one of a standard active neutrino.
The tightly coupled fluid will not affect structure formation in the same way as free-streaming neutrinos and the matter power spectrum does not deviate significantly from the prediction of a pure $\Lambda$CDM model. The residual difference in the matter power spectrum is not induced by the sterile neutrino mass, but instead by the increase in $\neff$, i.e.\ by the increase of the energy density of relativistic species, due to the presence of the additional neutrino and pseudoscalar degrees of freedom.

\subsection{Light sterile neutrinos in the pseudoscalar model}
\label{sec:pseudoCosmo}

We implemented the phenomenology of the pseudoscalar-sterile neutrino fluid in the Boltzmann solver CAMB~\cite{Lewis:1999bs} and we then fit the theoretical model to various cosmological data as in section~\ref{sec:lsn}.
The parameter space is the same as~\eqref{eq:cosmoParSpace}, but now the physical interpretation of $\neff$ is different. 
We now have $\neff = 3.046+N_{\rm fluid}$, where $N_{\rm fluid}$ encodes the sterile neutrinos and pseudoscalar degrees of freedom, which, by the time of photon decoupling, are collisional and not free-streaming.
Notice that the active neutrinos are kept fixed to the standard cosmological value $3.046$: the reason is that the pseudoscalar coupling is diagonal in mass basis, therefore the Standard Model neutrino mass eigenstates never recouple to the pseudoscalar after they decouple from the thermal bath at $T\sim 1$ MeV.

We point out that this consideration about the active states leads to a parametrisation which is somewhat different from the one we used in~\cite{Archidiacono:2015oma}, where we assumed flavour equilibration between the sterile and the active sector after neutrino decoupling. However, this flavour equilibration only occurs when the Standard Model mass eigenstates are also partially affected by interactions with pseudoscalars as the two plasmas will otherwise be fully disconnected.

\begin{figure*}
\begin{tabular}{cc}
\includegraphics*[width=0.48\linewidth]{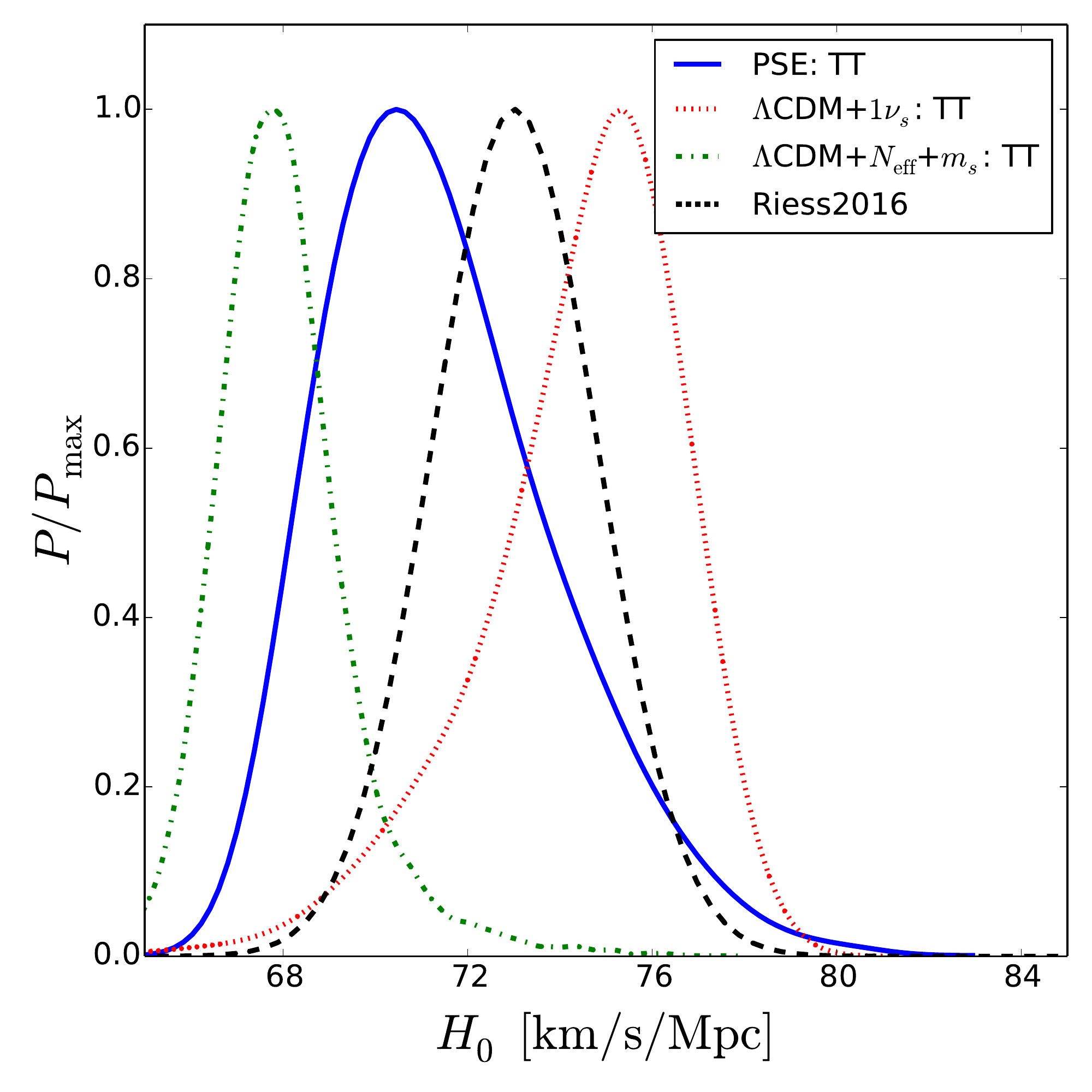}
&
\includegraphics*[width=0.48\linewidth]{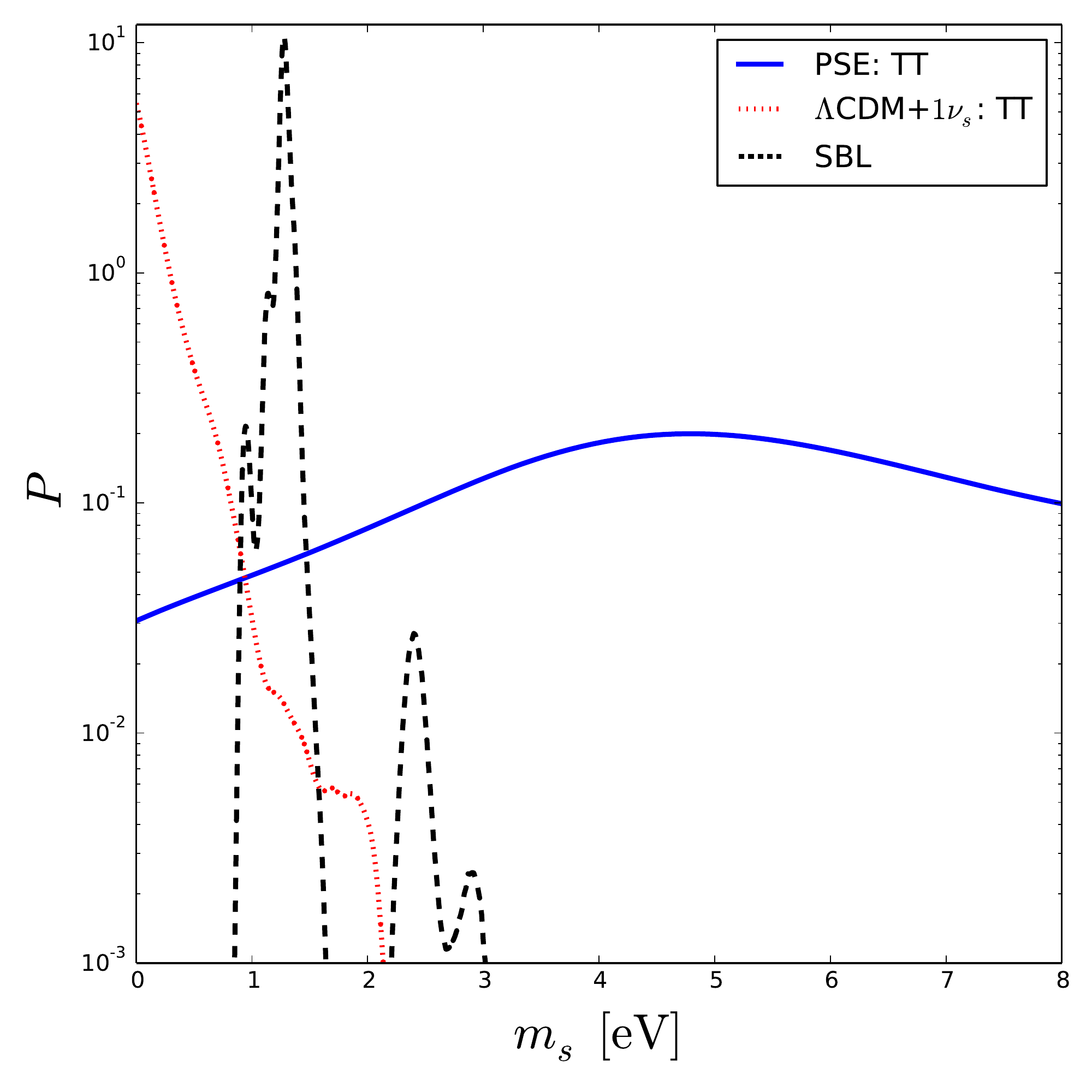}
\end{tabular}
\caption{ \label{fig:1dPost}
Comparison of the one-dimensional marginalised posterior distribution of the $H_0$ (left panel) and $m_s$ (right panel) parameters
as obtained from the analyses of the TT data in the
\lcdm+$1\nu_s$ model 
and in the pseudoscalar model.
In the left panel, we also report the local measure $H_0=73.00\pm1.75\,\,\Hou$~\cite{Riess:2016jrr} and the constraints obtained in the \lcdmnus\ model (see table~\ref{tab:lsnBounds}).
In the panel on the right, the posteriors are in logarithmic scale and are normalized such that they integrate to one. We also show the posterior for $m_s$ obtained in the SBL analysis described in section~\ref{sec:sbl}.
It is visible the secondary peak at $m_s\simeq2.4$, corresponding to the regions at $\Delta{m}^2_{41}\simeq6$ in figure~\ref{fig:sbl}.
}
\end{figure*}

As in section~\ref{sec:lsn}, $m_s$ is the sterile neutrino mass which in the pseudoscalar model determines the time when sterile neutrinos annihilate into pseudoscalars. In table \ref{tab:chi} we have listed the best-fit $\chi^2$ of the pseudoscalar model compared to the $\Lambda$CDM model for various data combinations. The preference for the pseudoscalar model is clearly strong, and even without including HST measurements it provides a better fit than the $\Lambda$CDM model. 
This is in contrast to models with a free $\neff$, where the $\chi^2_{\rm HST}$ gets better at the expenses of a worse $\chi^2_{\rm CMB}$.
In other words, the CMB $\chi^2$ is not affected by the inclusion of the prior on $H_0$.
The reason is that the pseudoscalar model naturally predicts a value of $H_0$ consistent with local measurements (see table~\ref{tab:constraints} and figure~\ref{fig:1dPost} left panel).
 
\begin{table*}[t]
\begin{center}
%\resizebox{0.5\textwidth}{!}{
\begin{tabular}{lccccl}
\hline
\hline
Data & $\chi^2_{\rm tot}$ & $\chi^2_{\rm CMB}$ & $\chi^2_{\rm HST}$ & $\chi^2_{\rm BAO}$ & Model\\[0.1cm]
\hline
 TT & $11260.5$ & $11260.5$ & $--$ & $--$ &  P \\ [0.1cm]
 & $11262.6$ & $11262.6$ & $--$ & $--$ &  $\Lambda$ \\ [0.1cm]
\hline
 TT+HST & $11260.2$ & $11260.2$ & $0.0$ & $--$ & P \\ [0.1cm]
 & $11272.1$ & $11265.6$ & $6.5$ & $--$ &   $\Lambda$\\ [0.1cm]
 \hline
 TT+BAO & $11267.6$ & $11263.2$ & $--$ & $4.4$ & P \\ [0.1cm]
 & $11270.5$ & $11265.6$ & $--$ & $4.9$ &   $\Lambda$\\ [0.1cm]
 \hline
TT+HST+BAO & $11269.1$ & $11262.8$ & $1.4$ & $4.9$ & P \\ [0.1cm]
& $11276.5$ & $11265.6$ & $6.5$ & $4.9$ &   $\Lambda$\\ [0.1cm]
\hline
TTTEEE & $12930.5$ & $12930.5$ & $--$ & $--$ &   P \\ [0.1cm]
& $12932.2$ & $12932.2$ & $--$ & $--$ &  $\Lambda$\\ [0.1cm]
\hline
\hline
\end{tabular}
%}
\end{center}
\caption{Best-fit $\chi^2$ for various models ($\Lambda$ stands for $\Lambda$CDM and $P$ for pseudoscalar model) and various data set combinations, as explained in the main text. The obtained values of $\chi^2$ are typically within $\Delta \chi^2 \sim 1$ of the true global best-fit value. 
}
\label{tab:chi}
\end{table*}

\begin{table*}[t]
\begin{center}
\resizebox{1\textwidth}{!}{
\begin{tabular}{lccccc}
\hline
\hline
 Parameter &TT&TT+HST&TT+BAO&TT+HST+BAO&TTTEEE\\[0.1cm]

\hline\noalign{\smallskip}
$H_0 \, [{\rm km/s/Mpc}]$& $71.4^{+1.8}_{-3.0}$&$72.4\pm2.5$&$69.8\pm1.4$&$71.1\pm1.2$&$70.9\pm1.8$\\[0.1cm]

\hline\noalign{\smallskip}
$\neff$&$<3.94$&$3.53\pm0.18$&$<3.67$&$3.49\pm0.18$&$<3.69$\\[0.1cm]

\hline
\hline
\end{tabular}
}
\end{center}
\caption{Constraints on $H_0$ and $\neff$ in the pseudoscalar model.
Marginalised constraints are given at $1\sigma$, while upper bounds are given at $2\sigma$. Data-set combinations are the same as in table~\ref{tab:chi}.}
\label{tab:constraints}
\end{table*}

Marginalised constraints on $H_0$ and $\neff$ are shown in table~\ref{tab:constraints}.
Including the $H_0$ prior leads to an evidence for a non-zero component of the pseudoscalar-sterile neutrino fluid at more than $2.5 \, \sigma$ ($3.1\,\sigma$ for TT+HST).
The constraints on $\neff$ are roughly the same in the pseudoscalar model as in the \lcdmnus\ 
model, but the difference in $\chi^2$-values are $\Delta \chi^2\sim - 5$,  i.e.\ there is a preference for additional degrees of freedom that are collisional/interacting rather than free-streaming.

\begin{figure}
\centering
\includegraphics[width=0.6\textwidth]{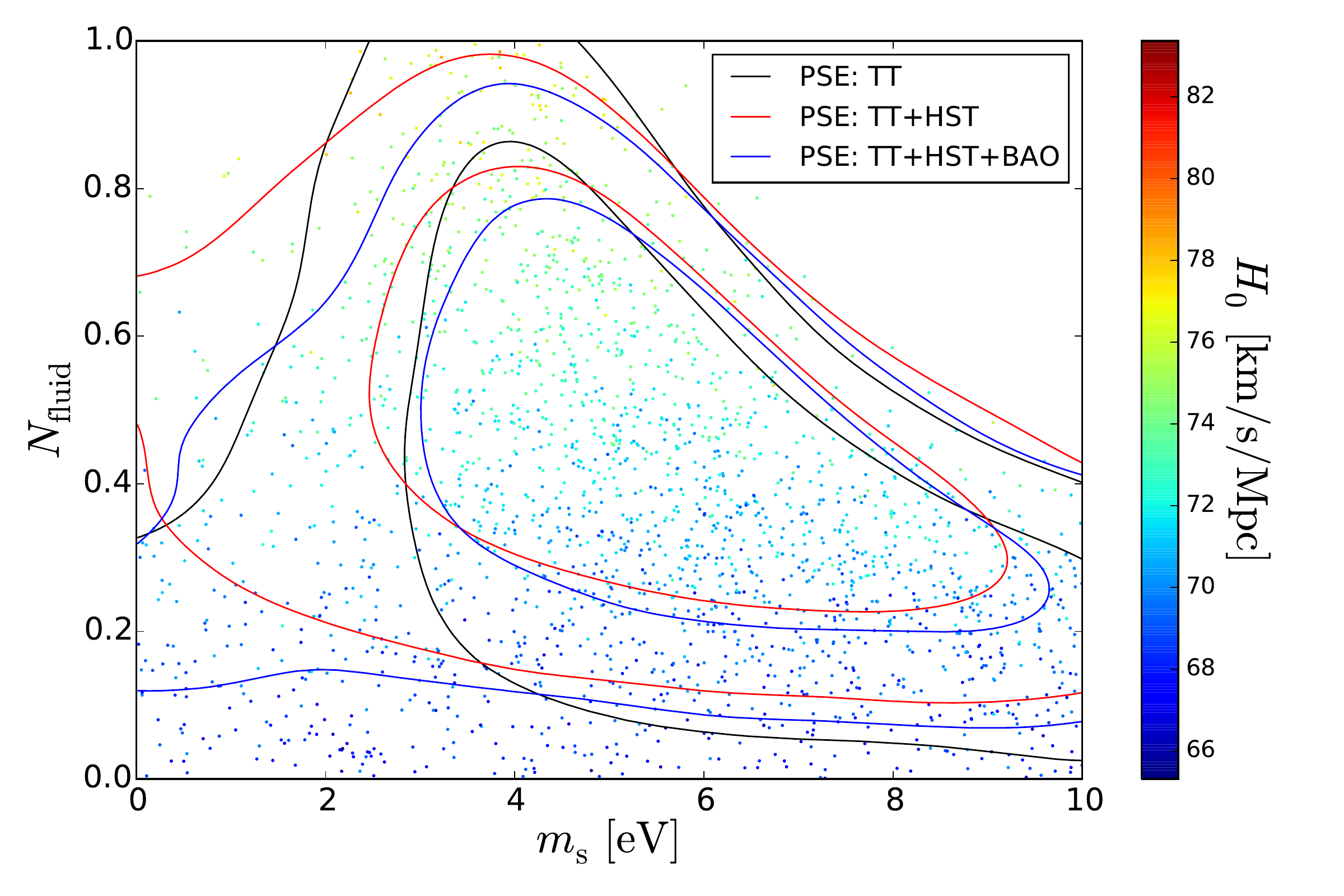}
\caption{Marginalised $1$ and $2\,\sigma$ contours in the plane $m_s$--$N_{\rm fluid}$ for the pseudoscalar model.
The points inside the contours are coloured according to the value of $H_0$ obtained by fitting only TT data.
}
\label{fig:nint}
\end{figure}

At the same time, the fact that the sterile neutrinos annihilate into
a massless pseudoscalar allows the bounds on $m_s$ to be less constraining.
In the right panel of figure~\ref{fig:1dPost} we present a comparison of
the constraints obtained from SBL analyses (black dashed line),
from the \lcdm+$1\nu_s$ model discussed in section~\ref{sec:lsn} (red dotted)
and from the pseudoscalar model (PSE, blue solid).
As we can see, in the pseudoscalar model the preferred mass is around
$m_s\simeq5$~eV and the posterior is wide enough to be compatible
with 1~eV.
For this reason, a combined fit of cosmological and SBL data is
possible.
Indeed, we will present the results of the joint analyses
in section~\ref{sec:joint}.
We also notice that in the pseudoscalar model $m_s$ and $N_{\rm fluid}$ are uncorrelated (see figure~\ref{fig:nint}); however, the allowed range of masses is larger when sterile neutrinos are partially thermalized.

\begin{figure}
\centering
\begin{tabular}{cc}
\includegraphics[width=0.5\textwidth]{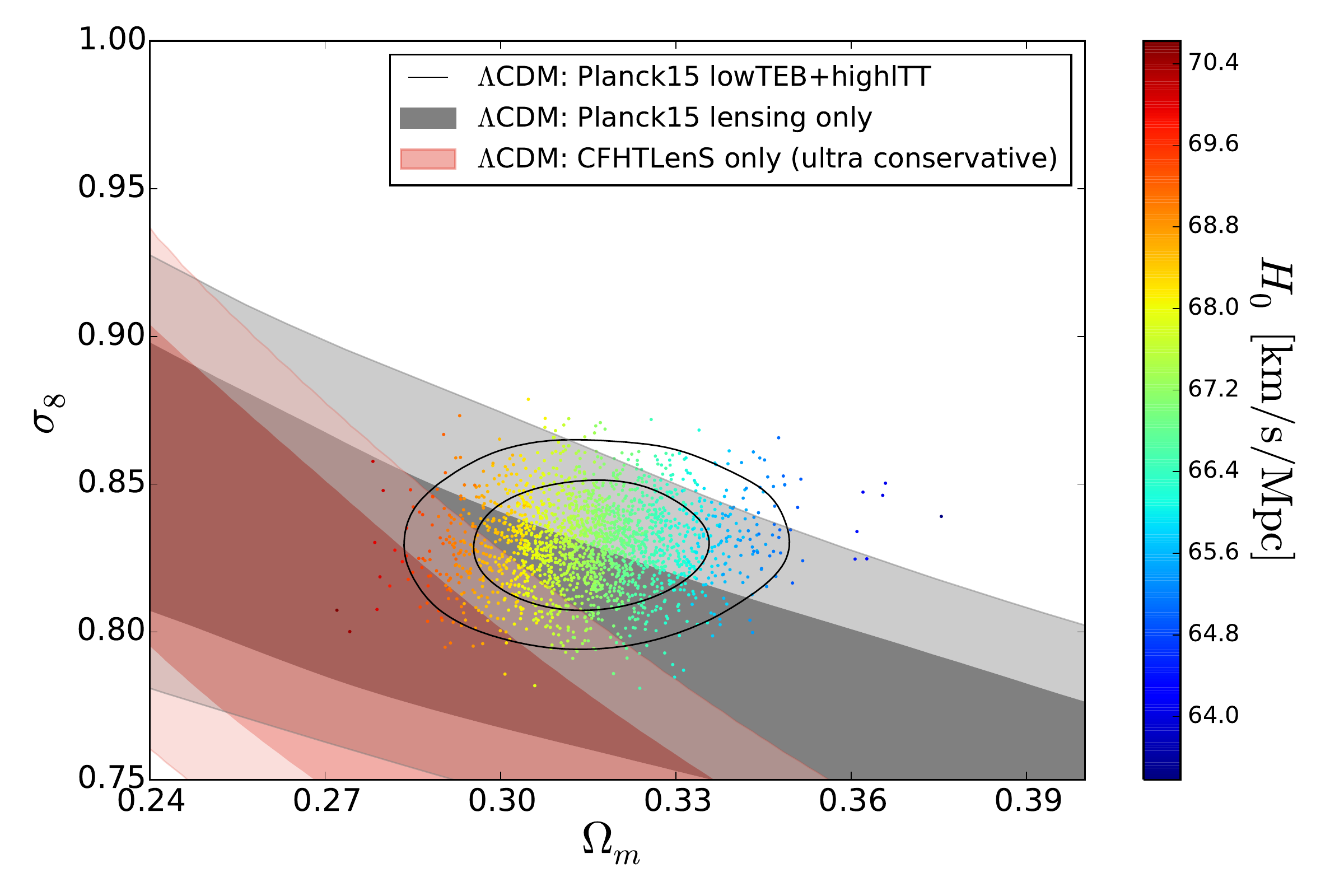}&
\includegraphics[width=0.5\textwidth]{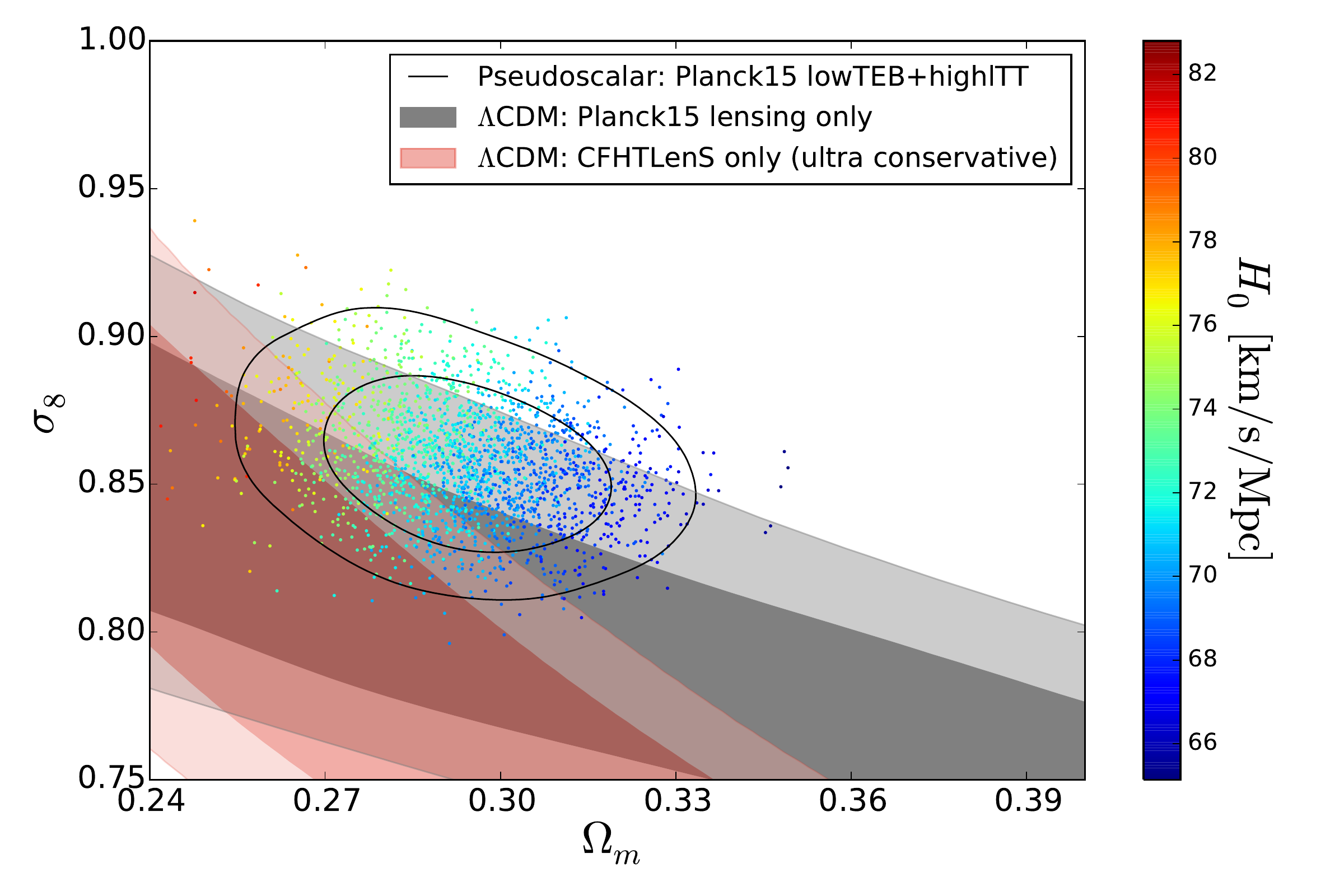}
\end{tabular}
\caption{Marginalised $1$ and $2\,\sigma$ contours in the plane $\sigma_8$--$\Omega_m$ of the $\Lambda$CDM model (left panel) and of the pseudoscalar model (right panel).
The points inside the contours are coloured according to the value of $H_0$ (see the scale on the right hand side of each plot).
The red and gray shaded areas show the constraints from Planck CMB lensing~\cite{Ade:2015zua}
and from CFHTLenS weak lensing data~\cite{Heymans:2013fya}, respectively.}
\label{fig:sigma8}
\end{figure}

Figure~\ref{fig:sigma8} shows that the consistency of the pseudoscalar model with the CFHTLenS weak lensing measurements is the same as in a pure $\Lambda$CDM model.
Indeed, weak gravitational lensing mainly probes a combination of $(H_0,\,\Omega_m,\,\sigma_8)$ on small scales.
Due to its peculiar late time phenomenology, the pseudoscalar model is consistent with local Universe measurements of $H_0$ and, at the same time, it does not induce any significant deviation of $\Omega_m$--$\sigma_8$ from the prediction of the $\Lambda$CDM model.
The preferred marginalised regions, however, slightly move in the direction of reconciling the cosmological predictions with the local determinations.

\section{Joint analyses}
\label{sec:joint}
In the following
we will present the results of the combination of the analyses of
short-baseline neutrino oscillation data
and of cosmological data
in the pseudoscalar model with two approaches:

\begin{itemize}

\item
In section~\ref{sec:cosmoSBL} we will analyse the cosmological data
using as a prior the marginal posterior probability for
$m_{s}=\sqrt{\Delta{m}^2_{41}}$
obtained from the short-baseline analysis.
This approach will allow us to get information on the cosmological parameters
taking into account the short-baseline data.

\item
In section~\ref{sec:implications} we will analyse the short-baseline neutrino oscillation data
using as a prior the marginal posterior probability for
$\sqrt{\Delta{m}^2_{41}}=m_{s}$
obtained from the analysis of cosmological data.
This approach will allow us to get information on the neutrino mixing parameters.

\end{itemize}

\subsection{SBL results as a prior in the cosmological analysis}
\label{sec:cosmoSBL}

\begin{figure*}
\begin{center}
\includegraphics*[width=0.48\linewidth]{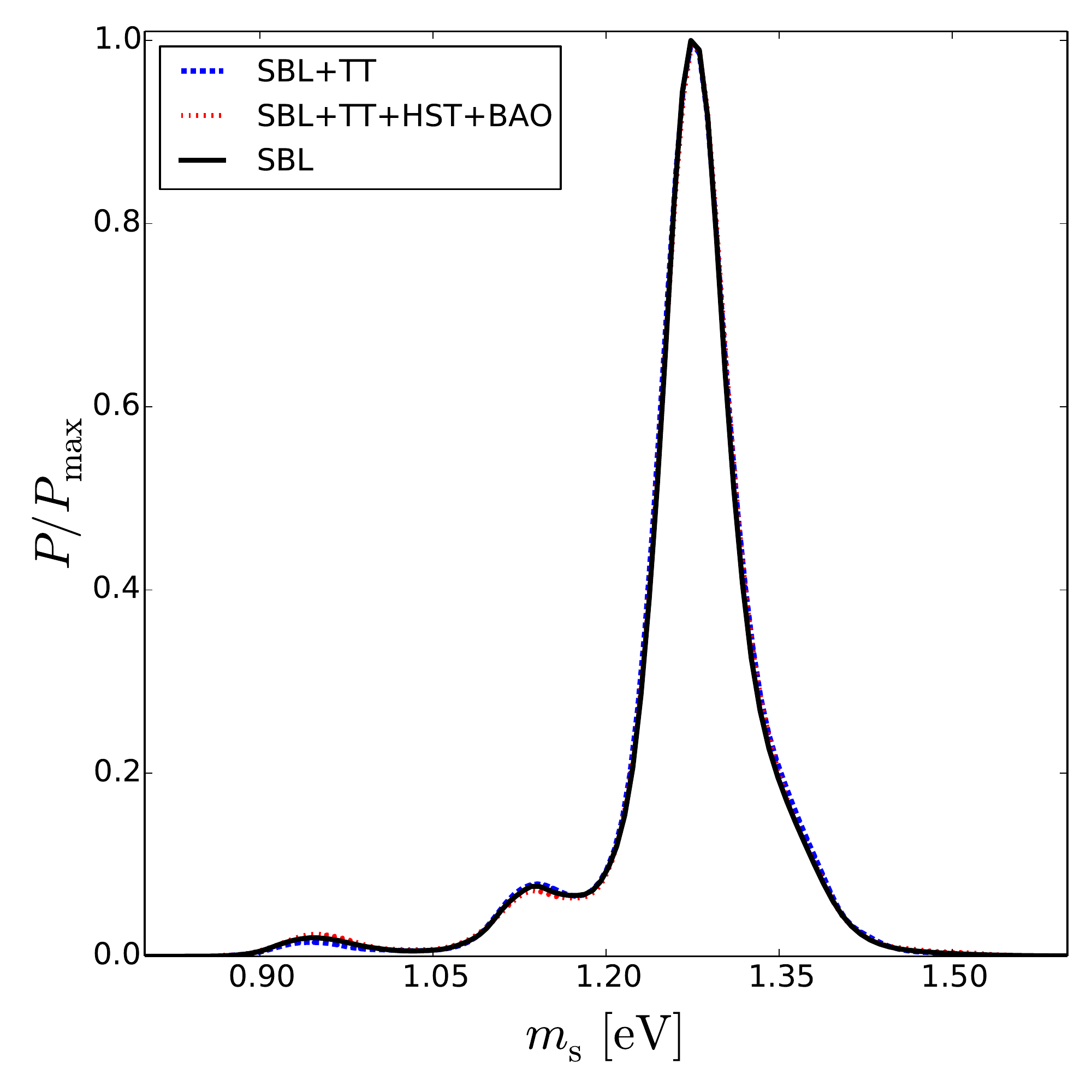}
\end{center}
\caption{
One-dimensional marginalised posterior distribution on the cosmological parameters $m_s$ obtained in the pseudoscalar model by including the SBL prior.
}
 \label{fig:1Dsbl}
\end{figure*}

As we have seen in section~\ref{sec:pseudoCosmo}, the pseudoscalar model on one hand provides a good fit to the cosmological data and on the other hand leads to cosmological results consistent with a light sterile neutrino. Therefore, it is timely to perform a combined analysis of cosmological data together with oscillation data in the framework of the pseudoscalar model. In order to do so we have followed the procedure described in section~\ref{sec:pseudoCosmo} and we add an extra $\chi^2$ obtained by fitting the cosmological sterile neutrino mass $m_s$ to the SBL prior described in section~\ref{sec:sbl}. As expected $\Delta \chi^2 \lesssim 1$, since the cosmological posterior on $m_s$ is broad and extends up to the oscillation preferred values of the sterile neutrino mass. Table~\ref{tab:sblconstraints} reports the constraints on $H_0$, $\neff$ and $m_s$. Given the consistency between oscillation and cosmological results, the constraints do not deviate significantly from the ones listed in table~\ref{tab:constraints} and the value of the Hubble constant is still in agreement with local Universe measurements. The main effect of adding the SBL prior is that the sterile neutrino mass is singled out around the oscillation best-fit, as shown in figure~\ref{fig:1Dsbl}.

\begin{table*}[t]
\begin{center}
%\resizebox{1\textwidth}{!}{
\begin{tabular}{lcccc}
\hline
\hline
 Parameter &SBL+TT&SBL+TT+HST&SBL+TT+BAO&SBL+TT+HST+BAO\\[0.1cm]

\hline\noalign{\smallskip}
$H_0 \, [{\rm km/s/Mpc}]$& $70.9^{+1.6}_{-3.3}$&$72.2\pm 1.7$&$69.19^{+0.83}_{-1.4}$&$70.6^{+1.1}_{-1.4}$\\[0.1cm]

\hline\noalign{\smallskip}
$\neff$&$<3.97$&$3.51\pm 0.20$&$<3.57$&$3.43^{+0.18}_{-0.22}$\\[0.1cm]

\hline\noalign{\smallskip}
$m_s$[eV]&$1.272^{+0.052}_{-0.038}$&$1.274^{+0.050}_{-0.038}$&$1.270^{+0.055}_{-0.035}$&$1.270^{+0.055}_{-0.035}$\\[0.1cm]

\hline
\hline
\end{tabular}
%}
\end{center}
\caption{Constraints on $H_0$, $\neff$ and $m_s$ in the pseudoscalar model for various cosmological data-set in combination with the SBL prior.
marginalised constraints are given at $1\sigma$, while upper bounds are given at $2\sigma$. Data-set combinations are the same as in table~\ref{tab:chi}.}
\label{tab:sblconstraints}
\end{table*}

\subsection{Cosmology as a prior in the SBL analysis}
\label{sec:implications}

\begin{figure*}
\begin{tabular}{cc}
\includegraphics*[width=0.48\linewidth]{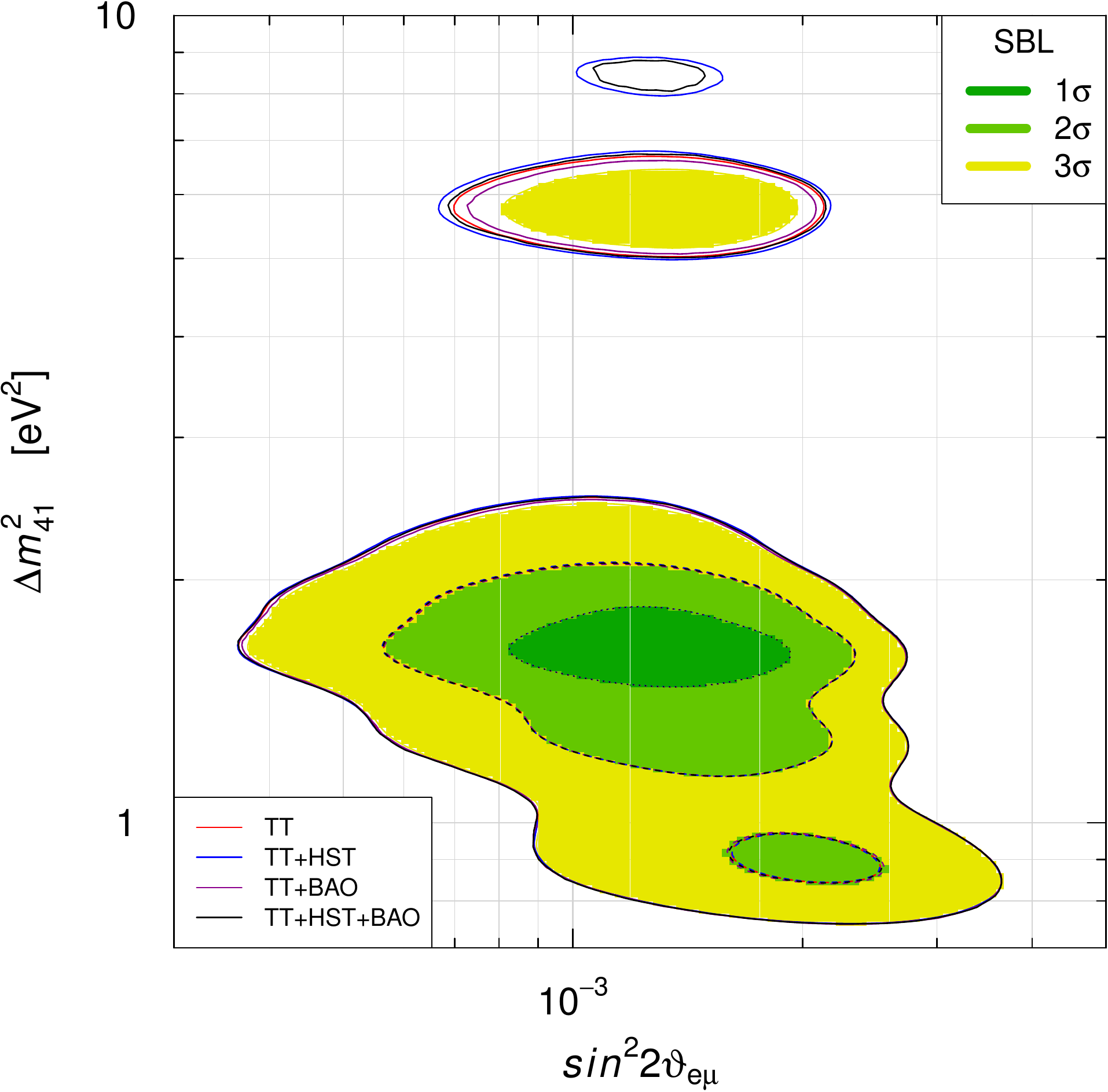}
&
\includegraphics*[width=0.48\linewidth]{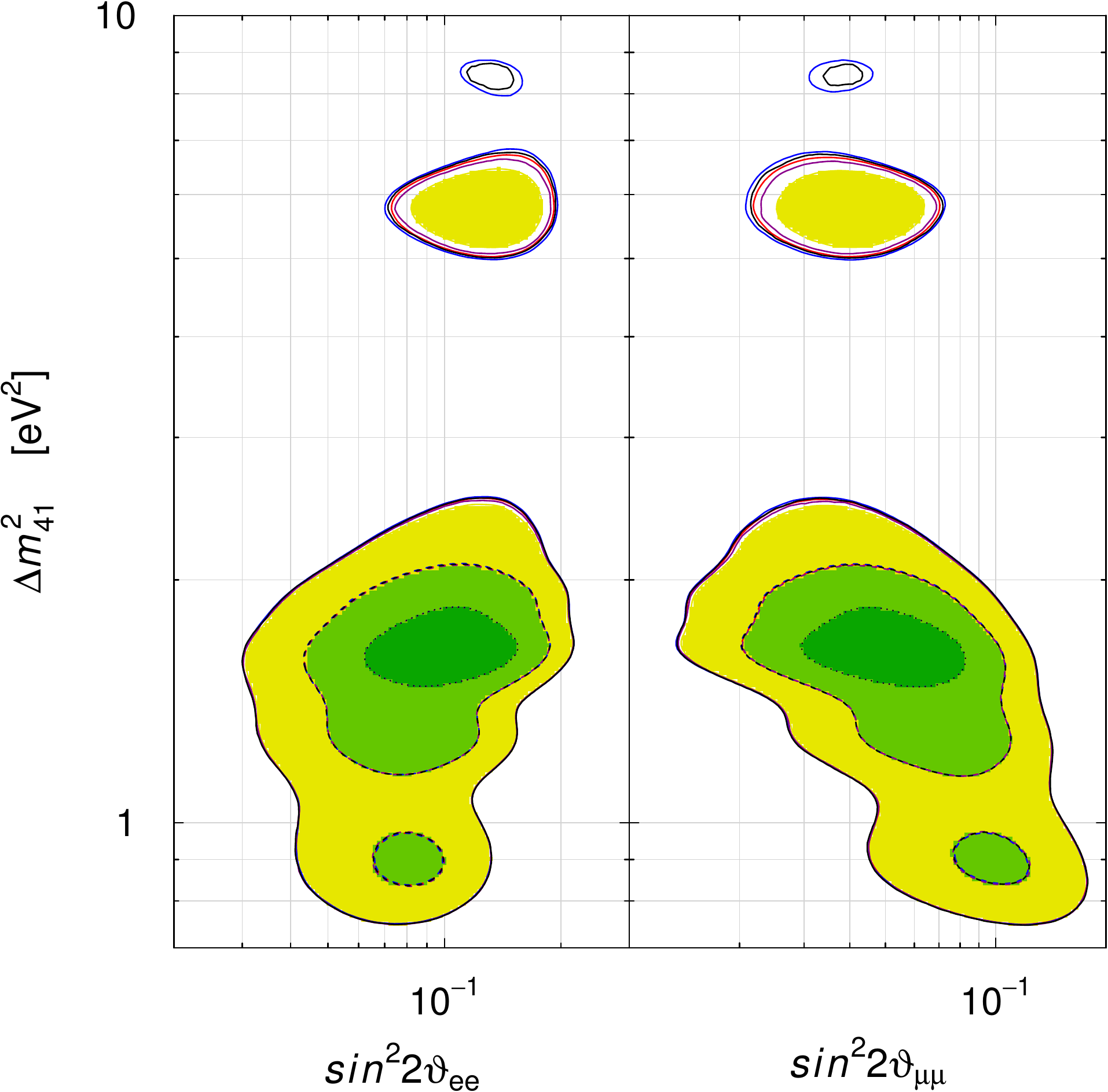}
\end{tabular}
\caption{ \label{fig:sbl-cosmo}
The coloured regions show the constraints in the planes of the effective amplitudes
$\sin^22\vartheta_{e\mu}$,
$\sin^22\vartheta_{ee}$ and
$\sin^22\vartheta_{\mu\mu}$
versus
$\Delta{m}^2_{41}$
which are allowed by the Bayesian global fit of short-baseline neutrino oscillation data, as in figure \ref{fig:sbl}.
The dotted, dashed and solid coloured curves enclose the regions allowed,
respectively, at
$1\sigma$,
$2\sigma$ and
$3\sigma$
in the combined analyses
with different priors on $\Delta{m}^2_{41}$
obtained from the analysis of cosmological data in the pseudoscalar model
using the
TT,
TT+HST,
TT+BAO and
TT+HST+BAO data sets (see section~\ref{sec:implications}).
A difference of the contours is visible only
for the large-$\Delta{m}^2_{41}$ regions.
}
\end{figure*}

It is interesting to investigate what are the implications of the analysis
of the
cosmological data in the pseudoscalar model
for the mixing parameters relevant for short-baseline neutrino oscillations.
We performed this study by considering the
posterior on $m_{s}$
obtained from the analysis of cosmological data
as a prior on
$\sqrt{\Delta{m}^2_{41}}=m_{s}$
in a Bayesian global analysis of
short-baseline neutrino oscillation data.
The results for the allowed regions in the
$\sin^22\vartheta_{e\mu}$--$\Delta{m}^2_{41}$,
$\sin^22\vartheta_{ee}$--$\Delta{m}^2_{41}$ and
$\sin^22\vartheta_{\mu\mu}$--$\Delta{m}^2_{41}$
planes
are shown in figure~\ref{fig:sbl-cosmo},
where we considered the
TT,
TT+HST,
TT+BAO and
TT+HST+BAO data sets.

From figure~\ref{fig:sbl-cosmo},
one can see that taking into account the cosmological data in the pseudoscalar model
has little effect on the values of the mixing parameters
that are favored by short-baseline data.
This is due to the compatibility of cosmological and short-baseline data
in the 3+1 pseudoscalar model already discussed in section~\ref{sec:pseudoCosmo}
and to the wide cosmological posteriors for $m_{s}$ in
the right panel of figure~\ref{fig:1dPost}.

The main small effect of cosmological data that one can notice
comparing the filled regions and the contour lines in figure~\ref{fig:sbl-cosmo}
is the increase of the Bayesian region allowed at $3\sigma$
at
$\Delta{m}^2_{41} \approx 6 \, \text{eV}^2$
and the emergence of a new Bayesian region allowed at $3\sigma$
at
$\Delta{m}^2_{41} \approx 8-9 \, \text{eV}^2$ in the fits which include the HST data.
These effects are
due to the preference for relatively large values of $m_{s}$
obtained in the analyses of cosmological data
and to the absence of strong constraints from
$\nua{e}$ and $\nua{\mu}$ disappearance
short-baseline experiments
at those values of $\Delta{m}^2_{41}$.
The indication in favor of values of $\Delta{m}^2_{41}$ larger than
about $1 \, \text{eV}^2$
is consistent with the recent constraints on short-baseline
$\nua{\mu}$ disappearance
with $\Delta{m}^2_{41} \lesssim 1 \, \text{eV}^2$
obtained in the IceCube~\cite{TheIceCube:2016oqi}
and MINOS~\cite{Timmons:2016hvv} experiments.

\section{Conclusions}
\label{sec:discussion}

We have performed an updated analysis of light sterile neutrinos in the context of cosmology.
A fourth, mainly sterile mass state with the mass and mixing needed to fit short baseline data is inevitably thermalised in the early Universe and its effect on structure formation and the CMB makes it extremely disfavoured by current data.
The fact that the presence of additional massless or massive neutrinos is disfavoured remains true even in extended cosmological models with significantly more parameters than the standard $\Lambda$CDM model~\cite{DiValentino:2016hlg}. 
However, if sterile neutrinos are charged under a new, non-standard interaction this picture can change dramatically.

Here we investigated how the model first suggested in~\cite{Archidiacono:2014nda} can alleviate the tension. The model is based on a interaction mediated by a light pseudoscalar particle, coupled only to the fourth neutrino mass eigenstate.
This model has a phenomenology which is very different from the standard picture.
First, the new interaction induces a large matter potential which can suppress equilibration of the additional mass state until after active neutrino decoupling.
Second, the interaction becomes very strong at late times and leads to a strongly coupled fluid of neutrinos and pseudoscalars.
This suppresses the anisotropic stress and surprisingly provides an excellent fit to all current data.
Very interestingly the model also predicts a value of the Hubble parameter much higher than the one obtained in the $\Lambda$CDM model and completely compatible with the locally measured value.

We found an excellent agreement of the cosmological posterior for the
sterile neutrino mass
$m_s$ with the squared-mass difference
$\Delta{m}^2_{41} \simeq m_s^2$
obtained from the analysis of short-baseline neutrino oscillation data.
The combined fit indicates a preference for values of $\Delta{m}^2_{41}$
larger than about $1 \, \text{eV}^2$,
in agreement with
the constraints obtained recently in the
IceCube~\cite{TheIceCube:2016oqi}
and MINOS~\cite{Timmons:2016hvv} experiments.

The question of the existence of the light sterile neutrino
indicated by the reactor, Gallium and LSND anomalies
will be answered in a definitive way in the next few years
through a wide program of many new neutrino oscillation experiments
which will investigate
the short-baseline disappearance of $\nua{e}$
produced in nuclear reactors and radioactive sources
and the short-baseline
$\nu_\mu \to \nu_e$ ($\bar{\nu}_\mu \to \bar{\nu}_e$)
transitions
and $\nua{\mu}$ 
disappearance of
accelerator neutrinos
(see the reviews in
refs.~\cite{Lasserre:2015eva,Lhuillier:2015fga,Caccianiga:2015ega,Spitz:2015gga,Gariazzo:2015rra,Giunti:2015wnd,
Stanco:2016gnl,Fava:2016vas}).
Moreover, in the near future the Euclid satellite will pin down the neutrino mass sum and the number of effective relativistic degrees of freedom \cite{Hannestad:2014voa}: if $\Sigma m_\nu$ is confirmed to be close to the minimum value of either the normal or inverted active neutrino hierarchy and, at the same time, there is an evidence for a value of $\neff$ slightly larger than the standard $3.046$, it will be a clear hint for physics beyond the standard model and beyond the standard picture of three massive neutrinos.
In other words, if, on one hand, neutrino experiments will confirm the existence of $\sim$~eV sterile neutrinos, and, on the other hand, precision cosmology will rule out additional neutrino-like species, though, indicating a deviation from $\neff=3.046$, the search for a new hidden sector, with new interactions and new particles, will be essential. In this scenario the pseudoscalar model would provide the simplest and most natural way of reconciling laboratory experiments with cosmological observations.

\section*{Note Added}

After the completion of this work the phenomenology of light sterile neutrinos
was discussed at the Neutrino 2016 Conference.
The Daya Bay and RENO Collaborations presented new upper limits on
$\sin^2 2\vartheta_{ee}$
for $3 \times 10^{-4} \lesssim \Delta{m}^2_{41} \lesssim 0.3 \, \text{eV}^2$,
which do not constrain the allowed region in Figs.~\ref{fig:sbl} and \ref{fig:sbl-cosmo}.
The MINOS Collaboration confirmed the upper limits on
$\sin^2 2\vartheta_{\mu\mu}$
for $3 \times 10^{-3} \lesssim \Delta{m}^2_{41} \lesssim 100 \, \text{eV}^2$
presented in ref.~\cite{Timmons:2016hvv},
that we discussed above.
The Daya Bay and MINOS results have been published too,
respectively,
in
ref.~\cite{An:2016luf}
and
ref.~\cite{MINOS:2016viw}.
Moreover,
the two collaborations presented a joint analysis of their data and
of the data of the Bugey-3 experiment \cite{Declais:1995su}
in order to constrain $\sin^2 2\vartheta_{e\mu}$
for $3 \times 10^{-4} \lesssim \Delta{m}^2_{41} \lesssim 100 \, \text{eV}^2$.
The results of this analysis,
also published in ref.~\cite{Adamson:2016jku},
are not relevant for our global fits,
because the data of Bugey-3 experiment are already taken into account
together with those of other reactor neutrino experiments
(see ref.~\cite{Gariazzo:2015rra}).

At the Neutrino 2016 Conference
there was also a presentation of the IceCube \cite{TheIceCube:2016oqi}
constraints on
$\sin^22\vartheta_{\mu\mu}$
for
$\Delta{m}^2_{41} \lesssim 1 \, \text{eV}^2$
discussed above.
These constraints have been taken into account in a global fit of short-baseline
neutrino oscillation data in ref.~\cite{Collin:2016aqd},
confirming the indication in favor of values of $\Delta{m}^2_{41}$ larger than
about $1 \, \text{eV}^2$
discussed in this paper.

\begin{acknowledgments}
S.G., C.G. and M.L. are supported by the research Grant {\sl Theoretical Astroparticle Physics} number 2012CPPYP7 
under the Program PRIN 2012 funded by the Ministero dell'Istruzione, Universit\`a e della Ricerca (MIUR).
R.H. acknowledges support from the Alexander von Humboldt Foundation.
S.H. acknowledges support from the Villum Foundation.
\end{acknowledgments}

% \newpage

% \bibliography{all}
% \bibliographystyle{JHEP}

\providecommand{\href}[2]{#2}\begingroup\raggedright\endgroup

\end{document}